\documentclass[12pt]{iopart}
\usepackage{iopams}
\usepackage{cite}
\usepackage{dsfont}
\usepackage{cancel}
\usepackage{graphicx}
\begin{document}

\title[Boltzmann magnetotransport theory with a nonvanishing Berry curvature]{Semiclassical Boltzmann magnetotransport theory in anisotropic systems with a nonvanishing Berry curvature}

\author{Jeonghyeon Suh$^{1,\dagger}$, Sanghyun Park$^{1,\dagger}$ and Hongki Min$^{1,*}$}

\address{$^1$ Department of Physics and Astronomy, Seoul National University, Seoul 08826, Korea}
\address{$^*$ Author to whom any correspondence should be addressed.}
\address{$^\dagger$ These authors contributed equally to this work.}
\ead{hmin@snu.ac.kr}

\begin{abstract}
Understanding the transport behavior of an electronic system under the influence of a magnetic field remains a key subject in condensed matter physics. Particularly in topological materials, their nonvanishing Berry curvature can lead to many interesting phenomena in magnetotransport owing to the coupling between the magnetic field and Berry curvature. By fully incorporating both the field-driven anisotropy and inherent anisotropy in the band dispersion, we study the semiclassical Boltzmann magnetotransport theory in topological materials with a nonvanishing Berry curvature. We show that as a solution to the Boltzmann transport equation the effective mean-free-path vector is given by the integral equation, including the effective velocity arising from the coupling between the magnetic field, Berry curvature and mobility. We also calculate the conductivity of Weyl semimetals with an isotropic energy dispersion, and find that the coupling between the magnetic field and Berry curvature induces anisotropy in the relaxation time, showing a substantial deviation from the result obtained assuming a constant relaxation time.
\end{abstract}

%
%
%

\section{Introduction}

The effect of a magnetic field on transport behavior has always been a topic of interest in condensed matter physics. Adding another tuning knob (magnetic field) to electronic transport experiments can enhance the understanding of the material of interest. In this regard, magnetotransport measurement can be a useful tool to reveal numerous fascinating features that a material hides. The quantum Hall effect \cite{Klitzing1980}, for example, has been brought to light by magnetoresistance (MR) experiments.

In particular, topological materials with a nonvanishing Berry curvature such as Weyl semimetals or topological insulators, display several interesting magnetotransport behaviors such as negative MR. Recently, the negative MR in Weyl semimetals \cite{Kim2013,Kim2014,Li2016,Zhang2016,Huang2015, Xiong2015, Li2015, Zhang2017, Li2016a, Arnold2016, Yang2015, Yang2015a, Wang2016, Zhang2017a,Nishihaya2018,Li2018,Wan2018} and topological insulators \cite{Wang2012,He2013,Wiedmann2016,Wang2015,Breunig2017,Assaf2017,Chen2020,Singh2018,Bhattacharyya2019,Dai2017, Andreev2018} has received significant attention. In a strong magnetic field regime, Landau-level-limited quantum magnetotransport is predominant \cite{Ishizuka2019, Lu2017, Behrends2017, Wang2018, Fu2020,Deng2019,Burkov2015, Zyuzin2012}, whereas in the weak magnetic field regime, the charge transport can be described by the semiclassical formalism \cite{Xiao2005,Xiao2010,Pal2010,Kim2013,Kim2014,Lundgren2014, Gao2017,Stephanov2012,Son2012,Son2013,Zyuzin2017,Sekine2017,Olson2007,Das2019,Chen2016,Dantas2018,Xiao2020}. 

Most of the studies using the semiclassical approach utilize a simple relaxation time approximation assuming isotropy of the system. However, this approximation could be problematic when the band dispersion of the system is highly anisotropic, and the system is no longer approximated as an isotropic system. Furthermore, in topological materials with a nonvanishing Berry curvature, this isotropic approximation cannot account for the anisotropy that arises from the coupling between the magnetic field and the Berry curvature. In other words, although the system is isotropic in the band dispersion, the magnetic field can induce anisotropy through coupling with the Berry curvature, which cannot be captured by the simple isotropic formalism.

With these motivations, a fully anisotropic Boltzmann magnetotransport equation is formulated that incorporates anisotropy from the energy dispersion as well as anisotropy arising from the coupling between the magnetic field and Berry curvature. Although several studies tried to consider a field-dependent anisotropy \cite{Li2018,Johansson2019}, our approach utilizes a more generally applicable method. In this work, the integral equation for the effective mean-free-path vector is obtained including the effective velocity due to the coupling between the magnetic field, Berry curvature and mobility. The Boltzmann equation is solved by introducing an ansatz for the nonequilibrium distribution function with a minimal set of assumptions that encompasses the electric field, magnetic field, and Berry curvature, and the corresponding magnetoconductivity was obtained. As an application of our method, we numerically calculate the conductivity of Weyl semimetals and find that the coupling between the magnetic field and Berry curvature is responsible for the field-driven anisotropy in the relaxation time, inducing a significant deviation from the result obtained assuming a constant relaxation time.

The rest of the paper is organized as follows. In \sref{sec:boltzmann_iso}, the Boltzmann magnetotransport equation for an isotropic system without a Berry curvature is summarized along with the demonstration of the relaxation time equation in electron gas systems. In \sref{sec:boltzmann_aniso}, the key results for the Boltzmann transport equation is presented that can be applied to a general system with anisotropy arising from the energy dispersion and the coupling between the magnetic field and the Berry curvature. In \sref{sec:conductivity}, the magnetoconductivity equations are given. In \sref{sec:numerical}, numerical calculations for the magnetoconductivity of Weyl semimetals are presented. Finally, the conclusions of the study are stated in \sref{sec:discussion}.

\section{Magnetotransport equation in electron gas systems}\label{sec:boltzmann_iso}
First, the magnetotransport relaxation time equation for an isotropic electron gas without a Berry curvature is introduced. In this case, the equation of motion for the Bloch electrons with charge $q$ under the influence of an electric field ${\bi E}$ and a magnetic field ${\bi B}$ takes a simple form \cite{Ashcroft1976}:
\begin{equation}
	\dot{\bi r} = {\bi v}_{\bi k}, \label{eq:rdot_3deg} \qquad
	\hbar \dot{\bi k} = q {\bi E} + \frac{q}{c} {\bi v}_{\bi k} \times {\bi B}, \label{eq:kdot_3deg}
\end{equation} 
where $\bi r$ is the position vector, $\bi k$ is the crystal momentum, ${\bi v}_{\bi k} = \frac{1}{\hbar}\nabla_{\bi k} \tilde{\varepsilon}_{\bi k}$, $\tilde{\varepsilon}_{\bi k} \equiv \varepsilon_{\bi k}-{\bi m}_{\bi k}\cdot {\bi B}$, $\varepsilon_{\bi k}= \frac{\hbar^2 k^2}{2 m}$ is the electronic band dispersion relation of an electron gas, $m$ is the effective electron mass, and ${\bi m}_{\bi k}$ is the orbital magnetic moment that vanishes for a single-band electron gas with no Berry curvature \cite{Xiao2005,Thonhauser2005,Ceresoli2006,Souza2008,Yao2008}.

The Boltzmann transport equation can be written as
\begin{equation}\label{eq:boltzmann}
	\frac{df}{dt}=\left(\frac{df}{dt}\right)_{\rm coll},
\end{equation}
where $f$ is the nonequilibrium distribution function, $\frac{df}{dt}$ is the time derivative of the distribution given by
\begin{equation}\label{eq:df_dt}
	\frac{df}{dt}=\frac{\partial f}{\partial t}+ \dot{\bi r} \cdot \nabla_{\bi r}f+\dot{\bi k} \cdot \nabla_{\bi k}f,
\end{equation}
and $\left(\frac{df}{dt}\right)_{\rm coll}$ is the collision integral term.

Now, it is assumed that $f$ is spatially homogeneous (in this case, no temperature gradient or nonuniform electric field) and has no explicit time dependence. Hence, \eref{eq:df_dt} becomes $\frac{df}{dt}=\dot{\bi k} \cdot \nabla_{\bi k} f$.
Note that the equilibrium Fermi-Dirac distribution function $f^{(0)}_{\bi k}$ depends on ${\bi k}$ via energy dispersion, that is, $f^{(0)}_{\bi k}= f^{(0)}(\tilde{\varepsilon}_{\bi k})$, the nonequilibrium distribution function $f_{\bi k}$ and its ${\bi k}$ gradient can be written as 
\begin{eqnarray}
	\nabla_{\bi k}f_{\bi k} 
	= \hbar {\bi v}_{\bi k} \frac{\partial f^{(0)}_{\bi k}}{\partial \tilde{\varepsilon}_{\bi k}} + \nabla_{\bi k} g_{\bi k},
\end{eqnarray}
where $g_{\bi k}=f_{\bi k}-f^{(0)}_{\bi k}$ is the part where the field-dependent terms are contained. Then to leading order in $\bi{E}$,
\begin{equation} \label{eq:eom}
	\frac{df}{dt} = \dot{\bi k} \cdot \nabla_{\bi k}f_{\bi k} \approx - q {\bi E}\cdot {\bi v}_{\bi k}S^{(0)}(\tilde{\varepsilon}_{\bi k})+\frac{q}{\hbar c} ({\bi v}_{\bi k} \times {\bi B}) \cdot\nabla_{\bi k} g_{\bi k},
\end{equation}
where $S^{(0)}(\tilde{\varepsilon}_{\bi k}) \equiv -\frac{\partial f^{(0)}_{\bi k}}{\partial \tilde{\varepsilon}_{\bi k}} $. 

The collision integral term $\left(\frac{df}{dt}\right)_{\rm coll}$ is given by \cite{Ashcroft1976}
\begin{equation}
	\left(\frac{df}{dt}\right)_{\rm coll} = - \int \frac{d^d k'}{(2\pi)^d} W_{\bi{k^\prime}\bi{k}}\left(g_{\bi k} - g_{\bi{k^\prime}}\right),
\end{equation}
where $d$ is the dimension of the system and $W_{\bi{k}'\bi{k}}=\frac{2\pi}{\hbar} n_{\rm imp} |V_{\bi{k}'\bi{k}}|^2 \delta(\tilde{\varepsilon}_{\bi{k}'}-\tilde{\varepsilon}_{\bi{k}})$ is the transition rate given by the Fermi's golden rule with the impurity potential $V_{\bi{k^\prime}\bi{k}}$ and the impurity density $n_{\rm imp}$. 

Thus, the Boltzmann equation becomes
\begin{equation}\label{eq:col_int}
	q {\bi E}\cdot {\bi v}_{{\bi k}} S^{(0)}(\tilde{\varepsilon}_{\bi k}) - \frac{q}{\hbar c}({\bi v}_{{\bi k}} \times {\bi B}) \cdot\nabla_{\bi k} g_{\bi k} = \int \frac{d^d k'}{(2\pi)^d} W_{\bi{k}'\bi{k}}\left(g_{\bi k} -g_{\bi{k^\prime}}\right).
\end{equation}
The definition and calculation of $g_{\bi k}$ is the primary concern for solving the Boltzmann equation. When there is no magnetic field, the simple relaxation time equation is often utilized, that is, $\left(\frac{df}{dt}\right)_{\rm coll}=-\frac{g_{\bi k}}{\tau_{\bi k}}$, assuming that the system relaxes back to equilibrium from the impurity scattering with a characteristic time scale given by the relaxation time $\tau_{\bi k}$. The relaxation time equation is extended to incorporate the additional contribution from the magnetic field, hence, the additional term $g_{\bi k}$ is given by
\begin{eqnarray}\label{eq:gk_isotropic}
	g_{\bi k} &=&q{\bi E} \cdot {\bi v}_{{\bi k}}  \tau_{\bi{k}} S^{(0)}(\tilde{\varepsilon}_{\bi k}) - \frac{q}{\hbar c} \tau_{\bi{k}} (\bi{v_k} \times {\bi B}) \cdot \nabla_{\bi k} g_{\bi k}.
\end{eqnarray}
Then, \eref{eq:gk_isotropic} can be iteratively expanded to 
\begin{equation}\label{eq:gk_iter_isotropic}
	g_{\bi{k}} = q\bi{E} \cdot \sum_{N=0}^{\infty} \hat{\mathds{S}}_{\bi{k}}^N \left[ \bi{v_k} \tau_{\bi{k}} S^{(0)} (\tilde{\varepsilon}_{\bi{k}}) \right],
\end{equation}
where $\hat{\mathds{S}}_{\bi{k}}$ is the operator defined by
\begin{equation}\label{eq:Dk_operator_isotropic}
	\hat{\mathds{S}}_{\bi{k}} \equiv - \frac{q}{\hbar c} \tau_{\bi{k}} (\bi{v_k} \times {\bi B}) \cdot \nabla_{\bi k}.
\end{equation}
It is assumed that $\tau_{\bi{k}}$ depends on $\bi{k}$ only through $\tilde{\varepsilon}_{\bi{k}}$. Since $\nabla_{\bi{k}} \tilde{\varepsilon}_{\bi{k}} = \hbar \bi{v_k}$, $\hat{\mathds{S}}_{\bi{k}}$ acting on $\tau_{\bi{k}}$ or $S^{(0)} (\tilde{\varepsilon}_{\bi{k}})$ vanishes. Thus, \eref{eq:gk_iter_isotropic} transforms into
\begin{equation}\label{eq:gk_isotropic_final}
	g_{\bi{k}} = \tau_{\bi{k}} \bi{v}_{\bi k}^{\rm eff} \cdot \bi{G}_{\bi k}^{(0)},
\end{equation}
where $\bi{G}_{\bi k}^{(0)} \equiv q\bi{E} S^{(0)}(\tilde{\varepsilon}_{\bi k})$ and $\bi{v}_{\bi k}^{\rm eff}$ is the effective velocity given by
\begin{equation}\label{eq:uk}
	\bi{v}_{\bi k}^{\rm eff} \equiv \sum_{N=0}^{\infty} \hat{\mathds{S}}_{\bi{k}}^N \bi{v_k} = \bi{v_k} + \sum_{N=0}^{\infty} \hat{\mathds{S}}_{\bi{k}}^N \left( \hat{\mathds{S}}_{\bi{k}} \bi{v_k} \right) = \bi{v_k} - \frac{\mu_{\bi{k}}}{c} \left( \bi{v}_{\bi k}^{\rm eff} \times \bi{B} \right),
\end{equation}
where $\mu_{\bi{k}} = {q\tau_{\bi{k}}}/{m}$ is the mobility of the electron gas system. Taking the vector product of $\frac{\mu_{\bi{k}}}{c} \bi{B}$ for each side of \eref{eq:uk},
\begin{equation}\label{eq:uk_inter}
	\frac{\mu_{\bi{k}}}{c} \bi{v}_{\bi k}^{\rm eff} \times \bi{B} = \frac{\mu_{\bi{k}}}{c} \bi{v_k} \times \bi{B} - \frac{\mu_{\bi{k}}^2}{c^2} \left[ \left( \bi{v}_{\bi k}^{\rm eff} \cdot \bi{B} \right) \bi{B} - B^2 \bi{v}_{\bi k}^{\rm eff} \right].
\end{equation}
Using $\frac{\mu_{\bi{k}}}{c} \bi{v}_{\bi k}^{\rm eff} \cdot \bi{B} = \frac{\mu_{\bi{k}}}{c} \bi{v_k} \cdot \bi{B}$ and \eref{eq:uk}, the following equation is obtained
\begin{equation}
	\bi{v}_{\bi k}^{\rm eff} - \bi{v_k} = -\frac{\mu_{\bi{k}}}{c} \bi{v_k} \times \bi{B} + \frac{\mu_{\bi{k}}^2}{c^2} \left[ \left( \bi{v_k} \cdot \bi{B} \right) \bi{B} - B^2 \bi{v}_{\bi k}^{\rm eff} \right].
\end{equation}
Thus, a closed form of $\bi{v}_{\bi k}^{\rm eff}$ is given by
\begin{equation}\label{eq:uk_closed}
	\bi{v}_{\bi k}^{\rm eff} = \frac{\bi{v_k} - \frac{\mu_{\bi{k}}}{c} \bi{v_k} \times \bi{B} + \frac{\mu_{\bi{k}}^2}{c^2} \left( \bi{v_k} \cdot \bi{B} \right) \bi{B}}{1 + \frac{\mu_{\bi{k}}^2}{c^2} B^2}.
\end{equation}
Note that from \eref{eq:uk}, the effective velocity can be alternatively written as
\begin{equation} \label{eq:v_eff_definition_alter_iso}
	\bi{v}_{\bi k}^{\rm eff} = \bi{v_k} \mathds{N}_{\bi{k}},
\end{equation}
where $\mathds{N}_{\bi{k}} \equiv \left( \mathds{1} - \frac{\mu_{\bi{k}}}{c} \mathds{F} \right)^{-1}$, $\mathds{F}$ is the field strength tensor defined by $\mathds{F}^{(ij)} = \sum_k \epsilon_{ijk} B^{(k)}$ \cite{Woo2022}, and $\epsilon_{ijk}$ is the Levi-Civita symbol.

Substituting $g_{\bi k}$ into \eref{eq:eom}, $\frac{df}{dt} = - \bi{v}_{\bi k}^{\rm eff} \cdot \bi{G}_{\bi k}^{(0)}$ is obtained. Then the Boltzmann equation can be expressed as
\begin{equation}\label{eq:col_int_2_add}
	\bi{v_k} \mathds{N}_{\bi{k}} \bi{G}_{\bi k}^{(0)} = \int \frac{d^d k'}{(2\pi)^d} W_{\bi{k^\prime}\bi{k}} \tau_{\bi{k}} \left( \bi{v_k} \mathds{N}_{\bi{k}} \bi{G}_{\bi k}^{(0)} - \bi{v_{k^\prime}} \mathds{N}_{\bi{k}}  \bi{G}_{\bi k}^{(0)} \right).
\end{equation}
Here, $\mathds{N}_{\bi{k}} = \mathds{N}_{\bi{k^\prime}}$ and $\bi{G}_{\bi k}^{(0)} = \bi{G}_{\bi k^\prime}^{(0)}$ were used since $\bi{G}_{\bi k}^{(0)}$ and $\mathds{N}_{\bi{k}}$ depend on ${\bi k}$ only through $\tilde{\varepsilon}_{\bi k}$. As $\bi{G}_{\bi k}^{(0)} \propto {\bi E}$ and \eref{eq:col_int_2_add} holds for all ${\bi E}$, after canceling out $\bi{G}_{\bi k}^{(0)}$ and multiplying $\mathds{N}_{\bi{k}}^{-1}$ on both sides the following can be obtained
\begin{eqnarray}
	\label{eq:relaxation_time_isotropic_magnetoelectric}
	\frac{1}{\tau_{\bi{k}}}&=& \int \frac{d^d k'}{(2\pi)^d} W_{\bi{k}'\bi{k}}\left(1-\cos\theta_{\bi{k}'\bi{k}} \right),
\end{eqnarray}
where $\theta_{\bi{k}'\bi{k}}$ is the angle between ${\bi k}$ and ${\bi k}'$. Note that \eref{eq:relaxation_time_isotropic_magnetoelectric} takes exactly the same form as 
the $\bi{B} = \boldsymbol{0}$ case
\cite{Ashcroft1976}. See \ref{sec:gk_isotropic_alter} for the review of the conventional derivation of the magnetotransport relaxation time equation for electron gas systems \cite{Ziman2001}.


In the presence of magnetic field, even in the systems with isotropic band dispersion, the relaxation time is generally expected to exhibit the anisotropic behavior due to the field-driven anisotropy. However for isotropic electron gas systems, the relaxation time is isotropic, that is, component-independent and depends on ${\bi k}$ only via energy, as seen in \eref{eq:relaxation_time_isotropic_magnetoelectric}. This interesting feature originates from the quadratic dispersion and the absence of the Berry curvature in the system. Since the corresponding mobility and $\mathds{N}_{\bi k}$ for the quadratic dispersion depend on ${\bi k}$ only via energy, $\mathds{N}_{\bi k}$ can be canceled off from the magnetotransport equation \eref{eq:col_int_2_add}, yielding the absence of preferred direction on the equation and thus the isotropy of the relaxation time. In contrast, if the system has anisotropic or non-quadratic dispersion, or a non-vanishing Berry curvature, the relaxation time is anisotropic in general so that the generalized magnetotransport equation in \sref{sec:boltzmann_aniso} must be used to correctly obtain the relaxation time.

\section{Magnetotransport equation in anisotropic systems with a nonvanishing Berry curvature}\label{sec:boltzmann_aniso}

Until this point, only an isotropic single-band system without a Berry curvature was considered, namely an isotropic electron gas. Upon removing this restriction, the anisotropy from the electronic band structure, as well as the anisotropy that arises from the external magnetic field coupled with the Berry curvature of the system can be accounted for. 

The semiclassical equation of motion for a Bloch electron in a system with a nonvanishing Berry curvature ${\bi \Omega}_{\bi{k}}$ is given by \cite{Sundaram1999}
\begin{equation}
	\label{eq:eom_berry}
	\dot{\bi r} = {\bi v}_{\bi{k}} - \dot{\bi k} \times {\bi \Omega}_{\bi{k}}, \qquad
	\hbar \dot{\bi k} = q {\bi E} + \frac{q}{c} \dot{\bi r} \times {\bi B}.
\end{equation}
Solving \eref{eq:eom_berry}, the following is obtained 
\numparts
	\label{eq:kdot_sol}
	\begin{eqnarray}
		\dot{\bi r} &=& \frac{1}{D_{\bi{k}}}\left[{\bi v}_{\bi{k}} - \frac{q}{\hbar} {\bi E} \times {\bi \Omega}_{\bi{k}}
		- \frac{q}{\hbar c} ({\bi v}_{\bi{k}} \cdot {\bi \Omega}_{\bi{k}}){\bi B} \label{eq:rdot_sol}\right], \\
		\hbar \dot{\bi k} &=& \frac{1}{D_{\bi{k}}}\left[q {\bi E}+ \frac{q}{c} {\bi v}_{\bi{k}} \times {\bi B}-\frac{q^2}{\hbar c}({\bi E} \cdot {\bi B}){\bi \Omega}_{\bi{k}}\right], 
	\end{eqnarray} 
\endnumparts
where $D_{\bi{k}} = 1 - \frac{q}{\hbar c} ({\bi \Omega}_{\bi{k}} \cdot {\bi B})$. In the presence of the Berry curvature, the phase-space volume element changes as $\Delta V_0 \to \Delta V_0/D_{\bi{k}}$ \cite{Xiao2005, Xiao2010}, modifying the density of states. Therefore, any integral over a Brillouin zone has an additional $D_{\bi{k}}$ factor to account for this change.

Then from \eref{eq:kdot_sol}, to the leading order in ${\bi E}$ we have
\begin{eqnarray} \label{eq:eom_2}
	\frac{df}{dt}&\approx&-\hbar \dot{\bi k} \cdot {\bi v}_{\bi{k}} S^{(0)}(\tilde{\varepsilon}_{\bi{k}}) + \dot{\bi k} \cdot\nabla_{\bi k} g_{\bi{k}} \nonumber \\
	&\approx&-\frac{1}{D_{\bi{k}}} \left[q {\bi E}-\frac{q^2}{\hbar c}({\bi E} \cdot {\bi B}){\bi \Omega}_{\bi{k}} \right] \cdot {\bi v}_{\bi{k}} S^{(0)}(\tilde{\varepsilon}_{\bi{k}}) + \frac{1}{D_{\bi{k}}} \frac{q}{\hbar c} ({\bi v}_{\bi{k}} \times {\bi B}) \cdot\nabla_{\bi k} g_{\bi{k}} \nonumber \\
	&=&-q{\bi E} \cdot {\bi v}^{\rm mod}_{\bi{k}} S^{(0)}(\tilde{\varepsilon}_{\bi{k}}) + \frac{q}{\hbar c}({\bi v}^{\rm mod}_{\bi{k}} \times {\bi B}) \cdot\nabla_{\bi k} g_{\bi{k}},
\end{eqnarray}
where ${\bi v}^{\rm mod}_{\bi{k}} = D_{\bi{k}}^{-1} \left[{\bi v}_{\bi{k}}-\frac{q}{\hbar c}({\bi \Omega}_{\bi{k}} \cdot {\bi v}_{\bi{k}}) {\bi B} \right]$
is the modified velocity coupled with the magnetic field and Berry curvature.

Then, the Boltzmann equation becomes
\begin{equation}\label{eq:col_int_3}
	\fl q {\bi E}\cdot {\bi v}^{\rm mod}_{\bi{k}} S^{(0)}(\tilde{\varepsilon}_{\bi{k}}) - \frac{q}{\hbar c}({\bi v}^{\rm mod}_{\bi{k}} \times {\bi B}) \cdot\nabla_{\bi k} g_{\bi{k}} = \int \frac{d^d k'}{(2\pi)^d} D_{\bi{k^\prime}} W_{\bi{k^\prime}\bi{k}}\left(g_{\bi{k}} -g_{\bi{k^\prime}}\right),
\end{equation}
which takes the form similar to (\ref{eq:col_int}), however, the velocity ${\bi v}_{\bi k}$ was replaced by the modified velocity ${\bi v}^{\rm mod}_{\bi{k}}$ with the additional $D_{\bi{k^\prime}}$ factor in the collision integral.

Extending (\ref{eq:gk_isotropic}) in the previous section, we introduce the modified ansatz for $g_{\bi k}$ as
\begin{eqnarray}\label{eq:gk_magneto_aniso}
	g_{\bi{k}} &=& q\bi{E} \cdot \bi{l}_{\bi{k}} S^{(0)} (\tilde{\varepsilon}_{\bi{k}}) - \frac{q}{\hbar c} \left( \bi{l}_{\bi{k}} \times \bi{B} \right) \cdot \nabla_{\bi{n}} g_{\bi{k}},
\end{eqnarray}
where $\bi{l}_{\bi{k}}$ is the mean-free-path vector analogous to $\bi{v_k} \tau_{\bi{k}}$ in \sref{sec:boltzmann_iso} and $\nabla_{\bi{n}} \equiv \nabla_{\bi{k}} - \hbar \bi{v}_{\bi{k}} \frac{\partial}{\partial \tilde{\varepsilon}_{\bi{k}}}$ is the surface gradient. A detailed justification for the ansatz is presented in \ref{sec:relaxation_time_approx_extension} and the consistency with the result in \sref{sec:boltzmann_iso} is discussed in \ref{sec:sec12_consistency}.
Unlike isotropic electron gas systems, $\bi{l}_{\bi{k}}$ is in general not parallel to $\bi{v}_{\bi{k}}$. Also note that the magnetic field couples only to the surface gradient of the distribution function since the Lorentz force does not affect the energy of the electrons.

Iteratively expanding \eref{eq:gk_magneto_aniso}, the following can be obtained
\begin{equation}\label{eq:gk_iter_aniso}
	g_{\bi{k}} = \bi{G}_{\bi{k}}^{(0)} \cdot \sum_{N=0}^{\infty} \hat{\mathds{S}}_{\bi{k}}^N \bi{l}_{\bi{k}} \equiv \bi{G}_{\bi{k}}^{(0)} \cdot \bi{L}_{\bi{k}},
\end{equation}
where $\bi{G}_{\bi{k}}^{(0)} \equiv q\bi{E} S^{(0)} (\tilde{\varepsilon}_{\bi{k}})$, $\hat{\mathds{S}}_{\bi{k}}$ is given by
\begin{equation}\label{eq:Dk_aniso}
	\hat{\mathds{S}}_{\bi{k}} \equiv -\frac{q}{\hbar c} \left( \bi{l}_{\bi{k}} \times \bi{B} \right) \cdot \nabla_{\bi{n}},
\end{equation}
and $\bi{L}_{\bi{k}}$ is the effective mean-free-path vector satisfying
\begin{equation}\label{eq:gk_self_aniso}
	\bi{L}_{\bi{k}} \equiv \sum_{N=0}^{\infty} \hat{\mathds{S}}_{\bi{k}} \bi{l}_{\bi{k}} = \bi{l}_{\bi{k}} + \hat{\mathds{S}}_{\bi{k}} \bi{L}_{\bi{k}},
\end{equation}
which describes the relation between the effective mean-free-path vector and conventional mean-free-path vector.

Reminding that $\bi{G}_{\bi{k}}^{(0)}$ depends on $\bi{k}$ only through $\tilde{\varepsilon}_{\bi{k}}$, the following integral equation for the effective mean-fee-path vector is obtained by putting $g_{\bi{k}}$ back into \eref{eq:col_int_3}
\begin{eqnarray}\label{eq:Lk_integral_equation}
	{\bi v}^{\rm mod}_{\bi{k}} - \frac{q}{\hbar c} \left[ ({\bi v}^{\rm mod}_{\bi{k}} \times {\bi B}) \cdot \nabla_{\bi n} \right] \bi{L}_{\bi{k}} = \int \frac{d^d k'}{(2\pi)^d} D_{\bi{k^\prime}} W_{\bi{k^\prime}\bi{k}} \left( \bi{L}_{\bi{k}} - \bi{L}_{\bi{k^\prime}} \right).
\end{eqnarray}
Here, $\bi{G}_{\bi{k}}^{(0)}$ was canceled out since $\bi{G}_{\bi{k}}^{(0)} \propto \bi{E}$ and the Boltzmann equation holds for all $\bi{E}$. Finding $\bi{L}_{\bi{k}}$ from \eref{eq:Lk_integral_equation}, the conductivity tensor of the systems would be obtained as discussed in \sref{sec:conductivity}. Note that from the charge conservation $\int \frac{d^d k}{(2\pi)^d} D_{\bi{k}} g_{\bi k} = 0$, we have the following constraint for $L_{\bi k}$:
\begin{equation}
	\label{eq:constraint_number_conservation}
	\int \frac{d^d k}{(2\pi)^d} D_{\bi{k}} \bi{L}_{\bi{k}} = 0.
\end{equation}

The alternative form of \eref{eq:Lk_integral_equation} would be useful for conceptual understanding. First, the effective velocity ${\bi v}^{\rm eff}_{{\bi k}}$ in \sref{sec:boltzmann_iso} can be generalized into
\begin{equation}\label{eq:v_eff_definition}
	{\bi v}^{\rm eff}_{\bi{k}} \equiv {\bi v}^{\rm mod}_{\bi{k}} - \frac{q}{\hbar c} \left[ ({\bi v}^{\rm mod}_{\bi{k}} \times {\bi B}) \cdot \nabla_{\bi n} \right] \bi{L}_{\bi{k}},
\end{equation}
so that \eref{eq:Lk_integral_equation} can be rewritten as
\begin{equation} \label{eq:Lk_integral_equation_alter}
	\bi{v}_{\bi k}^{{\rm eff}} = \int \frac{d^d k'}{(2\pi)^d} D_{{\bi k'}} W_{\bi{k^\prime}\bi{k}} \left( {\bi L_k} - {\bi L_{k^\prime}} \right).
\end{equation}
Notice that \eref{eq:Lk_integral_equation_alter} has exactly the same form as the anisotropic relaxation time equation at zero magnetic field
\begin{equation}
	\label{eq:integral_equation_nonmagnetic}
	\bi{v}_{\bi k} = \int \frac{d^d k'}{(2\pi)^d} W_{\bi{k^\prime}\bi{k}} \left( {\bi l_k} - {\bi l_{k^\prime}} \right),
\end{equation}
with the relaxation time $\tau_{\bi k}^{(i)} \equiv l_{\bi k}^{(i)} / v_{\bi k}^{(i)}$ \cite{Liu2016, Park2017, Park2019, Kim2019, Kawamura1992}, except that the velocity $\bi{v_k}$ and the mean-free-path vector $\bi{l_k}$ are replaced by the effective velocity $\bi{v}_{\bi k}^{\rm eff}$ and the effective mean-free-path vector ${\bi L_k}$, respectively, with the additional momentum-space volume factor $D_{{\bi k}}$ in the integral equation. On the other hand, the effective velocity given by \eref{eq:v_eff_definition} and the effective mean-free-path vector \eref{eq:gk_self_aniso} can be rewritten as
\begin{equation} \label{eq:v_eff_definition_alter}
	{\bi v}^{\rm eff}_{\bi{k}} = {\bi v}^{\rm mod}_{\bi{k}} \mathds{P}_{\bi{k}}, \qquad {\bi L}_{\bi k} = {\bi l}_{\bi k} \mathds{P}_{\bi{k}},
\end{equation}
where $\mathds{P}_{\bi{k}}$ is given by
\begin{equation}\label{eq:Pk_def}
	\mathds{P}_{\bi{k}}^{(ij)} \equiv \delta_{ij} + \frac{q}{\hbar c} \sum_k \mathds{F}^{(ik)} \nabla_{\bi{n}}^{(k)} L_{\bi{k}}^{(j)}.
\end{equation}
Substituting \eref{eq:v_eff_definition_alter} and iteratively solving \eref{eq:Pk_def}, the following equation is obtained
\begin{equation}\label{eq:Pk_alter}
	\mathds{P}_{\bi{k}} = \sum_{N=0}^{\infty} \left( \mathds{F} \frac{{\hat{\bmu}}_{\bi n}}{c} \right)^N \mathds{1} = \left( \mathds{1} - \mathds{F} \frac{{\hat{\bmu}}_{\bi n}}{c} \right)^{-1} \mathds{1},
\end{equation}
where ${\hat{\bmu}}_{\bi n}$ is the tensorial surface mobility operator defined by
\begin{equation}\label{eq:mu_n_def}
	{\hat{\bmu}}_{\bi n}^{(ij)} \mathds{Y}_{\bi{k}} = \frac{q}{\hbar} \nabla_{\bi{n}}^{(i)} \left( \bi{l_k} \mathds{Y}_{\bi{k}} \right)^{(j)}
\end{equation}
for an arbitrary matrix $\mathds{Y}_{\bi{k}}$. Note that $\mathds{P}_{\bi{k}}$ in \eref{eq:v_eff_definition_alter} describes a transformation of the modified velocity and the mean-free-path vector to the effective velocity and the effective mean-free-path vector, respectively, by the Lorentz force. Recall that a similar transformation appears in \sref{sec:boltzmann_iso} in the form of $\mathds{N}_{\bi{k}}$ in \eref{eq:v_eff_definition_alter_iso}. 

Finally, note that the relaxation time, which linearly relates the mean-free-path vector ${\bi l}_{\bi k}$ to the velocity in the absence of the magnetic field, is natural to be extended to the tensorial $\btau_{\bi k}$ defined by
\begin{equation}
	\label{eq:tau_definition}
	{\bi L}_{\bi k} = {\bi v}_{\bi k}^{\rm eff} \btau_{\bi k}.
\end{equation}

\section{Magnetoconductivity}\label{sec:conductivity}

Given $f_{\bi{k}}$, the current density ${\bi J}$ is given by
\begin{eqnarray}
	{\bi J} &=& g_{\rm s} q \int{\frac{d^d {\bi k}}{(2 \pi)^d} D_{{\bi k}} {\dot{\bi r}} f_{{\bi k}}} \\
	&=& g_{\rm s} q \int \frac{d^d {\bi k}}{(2 \pi)^d}\left[{\bi v}_{{\bi k}} - \frac{q}{\hbar} {\bi E} \times {\bi \Omega}_{{\bi k}}-\frac{q}{\hbar c} ({\bi v}_{{\bi k}} \cdot {\bi \Omega}_{{\bi k}}){\bi B}\right] \left( f^{(0)}_{{\bi k}} + g_{\bi{k}} \right), \nonumber
\end{eqnarray}
where $g_{\rm s}$ is the spin degeneracy factor. Working out each term up to leading order in ${\bi E}$, the following is obtained 
\begin{equation}
	{\bi J}\approx {\bi J}^{\rm AHE}+{\bi J}^{\rm CME}+{\bi J}^{\rm ext},
\end{equation}
where
\begin{equation}
	{\bi J}^{\rm AHE}=-\frac{g_{\rm s}q^2}{\hbar} \int \frac{d^d {\bi k}}{(2 \pi)^d}{\bi E} \times {\bi \Omega}_{{\bi k}} f^{(0)}_{{\bi k}}
\end{equation}
is the anomalous Hall effect (AHE) term,
\begin{equation}
	{\bi J}^{\rm CME}=-\frac{g_{\rm s}q^2}{\hbar c} \int \frac{d^d {\bi k}}{(2 \pi)^d} ({\bi v}_{{\bi k}} \cdot {\bi \Omega}_{{\bi k}}) {\bi B} f^{(0)}_{{\bi k}}
\end{equation}
is the chiral magnetic effect (CME) term, and
\begin{equation}
	{\bi J}^{\rm ext}=g_{\rm s} q \int \frac{d^d {\bi k}}{(2 \pi)^d} D_{{\bi k}}{\bi v}_{{\bi k}}^{\rm mod} g_{\bi{k}}
\end{equation}
is the extrinsic current term involving impurity scattering. 
Note that ${\bi v}_{{\bi k}} f^{(0)}_{{\bi k}}$ vanishes after integration.

The focus is on the extrinsic contribution and corresponding conductivity tensor $\sigma_{ij}^{\rm ext}$ defined by ${J}^{\rm ext}_i = \sum_{j} \sigma^{\rm ext}_{ij}E_{j}$.
For isotropic electron gas systems, from $g_{\bi k}$ given by \eref{eq:gk_isotropic_final} and ${\bi v}_{\bi k}^{\rm eff} = {\bi v}_{\bi k} \mathds{N}_{\bi k}$, the conductivity $\sigma_{ij}^{\rm ext}$ becomes
\begin{equation}\label{eq:conductivity_iso}
	\sigma_{ij}^{\rm ext} = g_{\rm s} q^2 \sum_l \int \frac{d^d {\bi k}}{(2 \pi)^d} S^{(0)} (\tilde{\varepsilon}_{\bi{k}}) v_{\bi k}^{(i)} v_{\bi{k}}^{(l)} \mathds{N}_{\bi k}^{(lj)} \tau_{\bi{k}}.
\end{equation}
For an anisotropic system with a nonvanishing Berry curvature, from \eref{eq:gk_iter_aniso} and \eref{eq:v_eff_definition_alter}, $\sigma_{ij}^{\rm ext}$ becomes
\begin{equation}\label{eq:conductivity_aniso}
	\sigma_{ij}^{\rm ext} = g_{\rm s} q^2 \int \frac{d^d {\bi k}}{(2 \pi)^d} D_{{\bi k}} S^{(0)} (\tilde{\varepsilon}_{\bi{k}}) v_{\bi{k}}^{{\rm mod}(i)} L_{\bi{k}}^{(j)},
\end{equation}
which reduces to \eref{eq:conductivity_iso} in isotropic electron gas systems with $\bi{v}_{\bi k}^{\rm mod} = \bi{v_k}$, $\bi{v_k} \mathds{P}_{\bi k} = \bi{v_k} \mathds{N}_{\bi k}$ (see \ref{sec:sec12_consistency}), and ${\bi L}_{{\bi k}} = \bi{v_k} \tau_{\bi k}$.

At zero magnetic field, $\mathds{P}_{\bi k} = \mathds{1}$, ${\bi v}_{\bi k}^{\rm mod} = {\bi v}_{\bi k}$, $\tilde{\varepsilon}_{\bi{k}} = \varepsilon_{\bi{k}}$, $D_{\bi k} = 1$, and $L_{\bi k}^{(i)} = v_{\bi k}^{(i)} \tau_{\bi k}^{(i)}$ \cite{Liu2016, Park2017, Park2019, Kim2019}; thus,
\begin{eqnarray}
	\sigma_{ij}^{\rm ext} = g_{\rm s} q^2  \int \frac{d^d {\bi k}}{(2 \pi)^d} S^{(0)} (\varepsilon_{\bi{k}}) v_{\bi k}^{(i)} v_{\bi k}^{(j)}\tau_{\bi{k}}^{(j)},
\end{eqnarray}
which is consistent with the conductivity equation obtained for an anisotropic system in the absence of a magnetic field \cite{Park2017, Park2019}.

\section{Numerical calculations for Weyl semimetals} \label{sec:numerical}
In this section, the relaxation time and the conductivity of Weyl semimetals are numerically calculated assuming the short-range impurities. In Weyl semimetals, the low-energy effective Hamiltonian for each node with chirality $\chi = \pm 1$ is given by $H_\chi (\bi{k}) = \chi \hbar v_0 \bsigma \cdot \bi{k}$ which has isotropic linear dispersion, where $\bsigma$ is the Pauli matrix vector. The magnetic field is assumed to be applied in the $z$ direction with ${\bi B} = B \hat{\bi z}$ ($B>0$). Note that the orbital magnetic moment and Zeeman splitting are neglected for simplicity, and the internode scattering is assumed to be negligible. Here we focus on the $xx$ component of the conductivity. For the $zz$ component along with the detailed analysis on the effect of the chiral anomaly, we leave it as a future work.

We obtain the effective mean-free-path vector ${\bi L}_{\bi k}$ by numerically solving \eref{eq:Lk_integral_equation}. Then, the relaxation time defined by \eref{eq:tau_definition} and the magnetoconductivity given by \eref{eq:conductivity_aniso} are directly obtained from ${\bi L}_{\bi k}$. For the details, see \ref{sec:numerical_calculation_detail}.

S. Woo \textit{et al.} \cite{Woo2022} introduced the dimensionless parameters $b_{{\rm BC}}$ and $b_\mu$ illustrating the coupling strength of the magnetic field with the Berry curvature and mobility, respectively, at the Fermi energy which are defined by 
\begin{equation}
	\label{eq:b_BC_def}
	b_{{\rm BC}} \equiv \frac{q}{\hbar c} \left| \bOmega_{\bi k} \right| \left| {\bi B} \right| = \frac{qB}{2\hbar c k_{\rm F}^2},
\end{equation}
and
\begin{equation}
	\label{eq:b_mu_def}
	b_\mu \equiv \frac{ \mu }{c} \left| {\bi B} \right| = \frac{qv_0 \tau_{\rm tr} B}{\hbar c k_{\rm F}},
\end{equation}
where $k_{\rm F}$ is the Fermi wavevector, and $\mu$ and $\tau_{\rm tr}$ are the mobility and relaxation time, respectively, at zero magnetic field. Note that the ratio $b_\mu / b_{{\rm BC}} \equiv 2k_{\rm F} l_{\rm F}$, where $l_{\rm F} \equiv v_0 \tau_{\rm tr}$ is the mean-free-path at the Fermi energy in the absence of the magnetic field.

\begin{figure}[h]
	\includegraphics[width=0.7\linewidth]{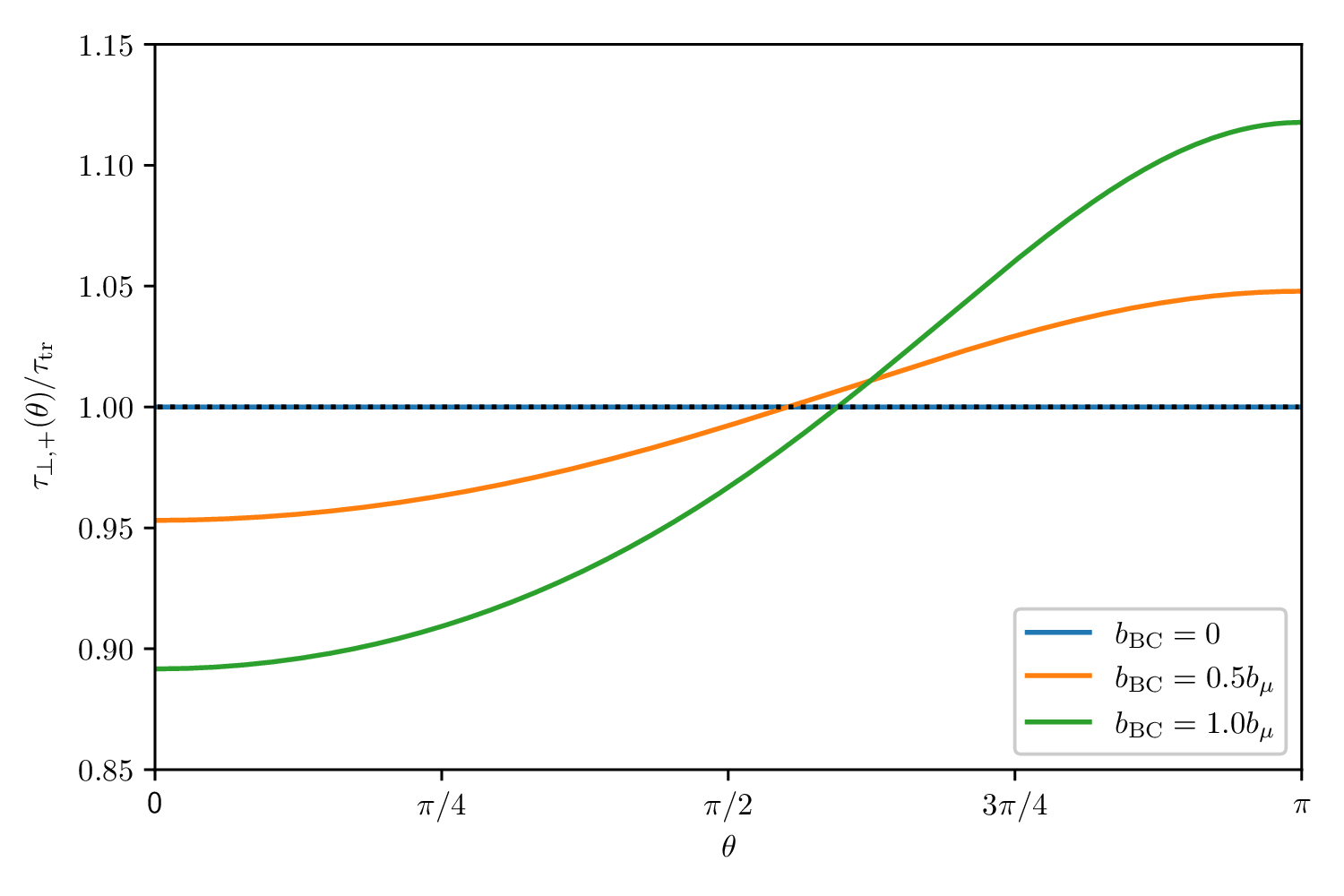}
	\centering
	\caption{
		$\tau_{\perp, +} (\theta)$ normalized by $\tau_{\rm tr}$ as a function of $\theta$ in the $\chi = +1$ node for $b_{\rm BC} / b_{\mu}$ = 0, 0.5, and 1, where $\tau_{\rm tr}$ is the relaxation time in the absence of the magnetic field. Here we set $b_\mu = 0.8$.
	} 
	\label{fig:tau_xy_plot}
\end{figure}
Using these parameters, figure \ref{fig:tau_xy_plot} illustrates the relaxation time $\tau_{\perp,+} (\theta)$ in the $xy$ plane (see \ref{sec:numerical_calculation_detail} for the detailed definition) for the $\chi = +1$ node as a function of $\theta$, the azimuthal angle of the momentum ${\bi k}$. From figure \ref{fig:tau_xy_plot}, we find that the anisotropy of the relaxation time induced by the magnetic field becomes larger as the coupling with the Berry curvature, $b_{\rm BC}$, increases.
Notice that when $b_{\rm BC} = 0$, the anisotropy in the relaxation time completely vanishes, yielding the same result with the one obtained from the conventional relaxation time approximation with a constant relaxation time \cite{Woo2022, Son2013}. This indicates that the coupling between the magnetic field and Berry curvature is responsible for the anisotropy in the relaxation time.

\begin{figure}[h]
	\includegraphics[width=0.7\linewidth]{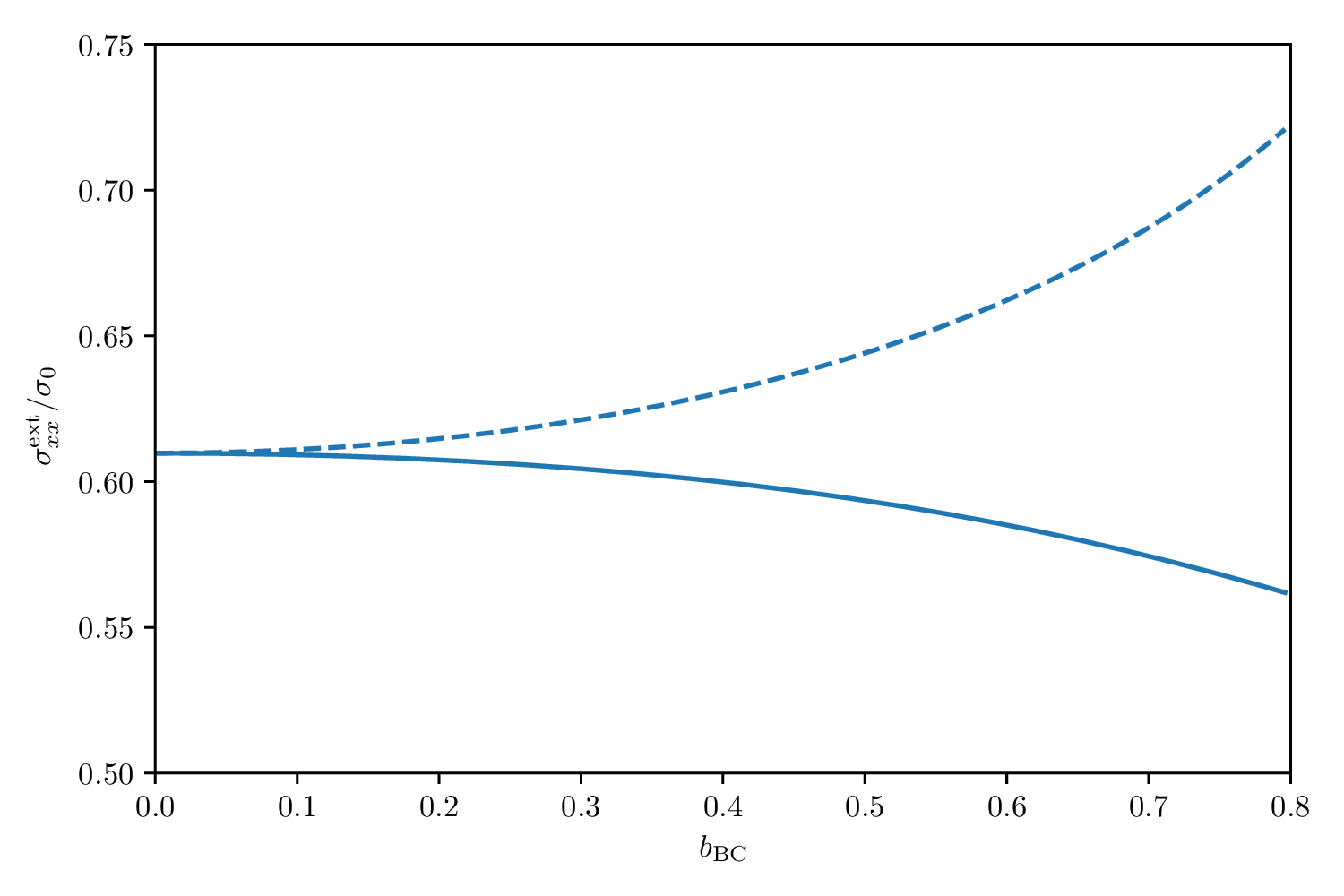}
	\centering
	\caption{
		The $xx$ component of the conductivity at zero temperature normalized by $\sigma_0$ as a function of $b_{\rm BC}$ for $b_\mu = 0.8$, where $\sigma_0$ is the conductivity in the absence of the magnetic field. The result obtained assuming a constant relaxation time is presented by the dashed lines for comparison.
	} 
	\label{fig:sigma_xx_plot}
\end{figure}

Figure \ref{fig:sigma_xx_plot} illustrates the $xx$ component of the conductivity. As the coupling between the magnetic field and Berry curvature increases, the result substantially deviates from that obtained assuming a constant relaxation time \cite{Woo2022}.

\section{Discussion} \label{sec:discussion}

In this study, the semiclassical magnetotransport equations as well as the corresponding mean-free-path equation were derived with a minimal set of assumptions imposed to obtain the compact closed form of the nonequilibrium distribution function. The field-dependent, anisotropic effective mean-free-path vector ${\bi L}_{\bi k}$ was obtained as a solution to the nonequilibrium distribution function, which is essential for understanding the effects of impurity scattering and the magnetic field on the transport behavior. We also numerically calculated the conductivity of Weyl semimetals and found that the coupling between the magnetic field and Berry curvature induces the anisotropy in the relaxation time, yielding a substantial deviation from the result obtained assuming a constant relaxation time.

As shown in \sref{sec:numerical} for Weyl semimetals, our extended Boltzmann transport theory can be applied not only when the system is inherently anisotropic, that is, the band dispersion is anisotropic, but also when the system is made to be anisotropic owing to the magnetic field. The coupling between the magnetic field and nonvanishing Berry curvature makes the distribution of electrons anisotropic, so the movement of electrons can no longer be described within an isotropic formalism, even when the band dispersion of the system is isotropic. Instead, the (effective) mean-free-path is given by the coupled integral equation as the band velocity ${\bi v}_{\bi k}$ acquires an additional contribution arising from the magnetic field coupled with the Berry curvature and mobility. Using the proposed formalism, any anisotropy of the system can be properly assessed, regardless of its origin. We stress that previous studies on the magnetotransport did not consider the Berry-curvature-induced anisotropy, which is essential to correctly describe transport in the systems with a nonvanishing Berry curvature.

\ack

This work was supported by the National Research Foundation of Korea (NRF) grant funded by the Korea government (MSIT) (No. 2018R1A2B6007837) and Creative-Pioneering Researchers Program through Seoul National University (SNU).

\appendix

\section{Conventional derivation of the magnetotransport equation in electron gas systems} \label{sec:gk_isotropic_alter}
In this section, we review the conventional derivation of the magnetotransport relaxation time equation in electron gas systems in \cite{Ziman2001} with some revision on mathematically unrigorous parts.

The alternative form of \eref{eq:eom} and \eref{eq:gk_isotropic} can be written as $\frac{df}{dt} = -\bi{v_k} \cdot \bi{G_k}$ and $g_{\bi{k}} = \tau_{\bi{k}} \bi{v_k} \cdot \bi{G_k}$, where 
\begin{equation} \label{eq:Gk_isotropic_def}
	\bi{G_k} \equiv \bi{G}_{\bi k}^{(0)} + \frac{q}{\hbar c} \left( \nabla_{\bi{k}} g_{\bi{k}} - A_{\bi{k}} \bi{v_k} \right) \times \bi{B}.
\end{equation}
Here $A_{\bi{k}}$ is chosen so that $\bi{G_k}$ depends on $\bi{k}$ only via $\tilde{\varepsilon}_{\bi{k}}$.
Substituting $g_{\bi{k}}$ into \eref{eq:Gk_isotropic_def}, the following is obtained
\begin{equation} \label{eq:Gk_isotropic}
	\bi{G_k} \equiv \bi{G}_{\bi k}^{(0)} +  \frac{\mu_{\bi{k}}}{c} \bi{G_k} \times \bi{B},
\end{equation}
where $\mu_{\bi{k}} = q\tau_{\bi{k}} / m$ is the mobility and $A_{\bi{k}}$ is given by
\begin{equation}
	A_{\bi{k}} = \hbar\bi{v_k} \cdot \frac{\partial}{\partial \tilde{\varepsilon}_{\bi{k}}} ( \tau_{\bi{k}} \bi{G_k} ).
\end{equation}
Note that we used $\bi{v_k} = \hbar \bi{k} / m$ in the electron gas systems. Taking the vector product of $\frac{\mu_{\bi{k}}}{c} \bi{B}$ for each side of \eref{eq:Gk_isotropic},
\begin{equation}\label{eq:Gk_inter}
	\frac{\mu_{\bi{k}}}{c} \bi{G_k} \times \bi{B} = \frac{\mu_{\bi{k}}}{c} \bi{G}_{\bi k}^{(0)} \times \bi{B} + \frac{\mu_{\bi{k}}^2}{c^2} \left[ \left( \bi{G_k} \cdot \bi{B} \right) \bi{B} - B^2 \bi{G_k} \right].
\end{equation}
Using $\frac{\mu_{\bi{k}}}{c} \bi{G_k} \cdot \bi{B} = \frac{\mu_{\bi{k}}}{c} \bi{G}_{\bi k}^{(0)} \cdot \bi{B}$ and \eref{eq:Gk_isotropic}, the following equation is obtained
\begin{equation}
	\bi{G_k} - \bi{G}_{\bi k}^{(0)} = \frac{\mu_{\bi{k}}}{c} \bi{G}_{\bi k}^{(0)} \times \bi{B} + \frac{\mu_{\bi{k}}^2}{c^2} \left[ \left( \bi{G}_{\bi k}^{(0)} \cdot \bi{B} \right) \bi{B} - B^2 \bi{G_k} \right].
\end{equation}
Thus, a closed form of $\bi{G_k}$ is given by
\begin{equation}\label{eq:Gk_closed}
	\bi{G_k} = \frac{\bi{G}_{\bi k}^{(0)} + \frac{\mu_{\bi{k}}}{c} \bi{G}_{\bi k}^{(0)} \times \bi{B} + \frac{\mu_{\bi{k}}^2}{c^2} \left( \bi{G}_{\bi k}^{(0)} \cdot \bi{B} \right) \bi{B}}{1 + \frac{\mu_{\bi{k}}^2}{c^2} B^2},
\end{equation}
which is consistent with the result in \cite{Ziman2001}, or equivalently $\bi{G_k} = \mathds{N}_{\bi{k}} \bi{G}_{\bi k}^{(0)}$. 

Substituting $\frac{df}{dt}$ and $g_{\bi{k}}$ to the Boltzmann equation, the following is obtained
\begin{equation}\label{eq:col_int_4}
	\bi{v_k}\cdot \bi{G_k} = \int \frac{d^d k'}{(2\pi)^d} W_{\bi{k}'\bi{k}} \tau_{\bi{k}} \left( \bi{v_k}\cdot \bi{G_k} - \bi{v_{k^\prime}}\cdot \bi{G_k} \right),
\end{equation}
where $\bi{G_k} = \bi{G_{k^\prime}}$ was used since $\bi{G_k}$ depends on $\bi{k}$ only via $\tilde{\varepsilon}_{\bi{k}}$ as can be verified in \eref{eq:Gk_closed}. As $\bi{G}_{\bi k}^{(0)} \propto \bi{E}$ and the Boltzmann equation holds for all $\bi{E}$, \eref{eq:col_int_4} reduces to \eref{eq:relaxation_time_isotropic_magnetoelectric} after canceling out $\bi{G}_{\bi k}^{(0)}$ and multiplying $\mathds{N}_{\bi{k}}^{-1}$ on both sides.

\section{Justification for the ansatz for $g_{\bi k}$}

\label{sec:relaxation_time_approx_extension}
Separating $\nabla_{\bi{k}} = \nabla_{\bi{n}} + \hbar \bi{v_k} \frac{\partial}{\partial \tilde{\varepsilon}_{\bi{k}}}$ and using the definition of $\bi{v}_{\bi k}^{\rm mod}$, \eref{eq:eom_2} becomes
\begin{eqnarray}\label{eq:eom_2_mod}
	\fl \frac{df}{dt} \approx -q{\bi E} \cdot {\bi v}^{\rm mod}_{\bi k} S^{(0)}(\tilde{\varepsilon}_{\bi k}) + \frac{q}{\hbar c}({\bi v}^{\rm mod}_{\bi k} \times {\bi B}) \cdot \nabla_{\bi n} g_{\bi k} + \cancel{\frac{q}{c} D_{\bi{k}} \left( {\bi v}^{\rm mod}_{\bi k} \times \bi{B} \right) \cdot {\bi v}^{\rm mod}_{\bi k} \frac{\partial g_{\bi{k}}}{\partial \tilde{\varepsilon}_{\bi k}}}.
\end{eqnarray}
Note that the last term on the right-hand side of \eref{eq:eom_2_mod} actually vanishes. To extend the conventional relaxation time approximation $\frac{df}{dt}=-\frac{g_{\bi{k}}}{\tau_{\bi{k}}}$, we need to focus on the correspondence between the Boltzmann transport theory and many-body diagrammatic theory. In the many-body diagrammatic theory, the relaxation time is introduced from the vertex correction to the current-current response function \cite{Kim2019, Flensberg2004, Coleman2016} transforming velocity ${\bi v}_{\bi k}$ into the current vertex $\Lambda^{(i)} (\bi k) = v_{\bi k}^{(i)} \tau_{\bi k}^{(i)} / \tau_{\bi k}^{\rm qp}$ \cite{Kim2019}, where $\tau_{\bi k}^{\rm qp}$ is the quasiparticle lifetime given by
\begin{equation}
	\frac{1}{\tau_{\bi k}^{\rm qp}} = \int \frac{d^d k'}{(2\pi)^d} W_{\bi{k}'\bi{k}}.
\end{equation}
Thus, it is reasonable to obtain $g_{\bi{k}}$ by replacing the (modified) velocity in $-\frac{df}{dt}$ with the mean-free-path vector. Thus, from \eref{eq:eom_2_mod}, we make an ansatz for $g_{\bi{k}}$ as
\begin{eqnarray}
	\label{eq:gk_magneto_aniso_ansatz}
	g_{\bi{k}} &=& q{\bi E} \cdot \bi{l_k} S^{(0)} (\tilde{\varepsilon}_{\bi k}) - \frac{q}{\hbar c}(\bi{l_k} \times {\bi B}) \cdot \nabla_{\bi n} g_{\bi k} \\ && - \cancel{\frac{q}{2c} D_{\bi{k}} \left[ \left( \bi{l_k} \times \bi{B} \right) \cdot {\bi v}^{\rm mod}_{\bi k} + \left( {\bi v}^{\rm mod}_{\bi k} \times \bi{B} \right) \cdot \bi{l_k} \right] \frac{\partial g_{\bi{k}}}{\partial \tilde{\varepsilon}_{\bi k}}}, \nonumber
\end{eqnarray}
which reduces to the suggested ansatz in \eref{eq:gk_magneto_aniso} since the last term vanishes. For the last term on the right-hand side of \eref{eq:gk_magneto_aniso_ansatz} where two modified velocities appear, we took the average after separately 
replacing one of ${\bi v}^{\rm mod}_{\bi k}$ with $\bi{l_k}$.
This choice is reasonable in that the vanishing term in \eref{eq:gk_magneto_aniso_ansatz} also originates from the vanishing term in \eref{eq:eom_2_mod}.

For isotropic electron gas systems where ${\bi v}^{\rm mod}_{\bi k} = \bi{v_k}$, $\bi{l_k} = \bi{v_k} \tau_{\bi{k}}$ and $D_{\bi{k}} = 1$, \eref{eq:gk_magneto_aniso_ansatz} becomes
\begin{equation}
	\label{eq:gk_magneto_iso_ansatz}
	\fl g_{\bi{k}} = q{\bi E} \cdot \bi{v_k} \tau_{\bi{k}} S^{(0)} (\tilde{\varepsilon}_{\bi k}) - \frac{q}{\hbar c} \tau_{\bi{k}} (\bi{v_k} \times {\bi B}) \cdot \nabla_{\bi n} g_{\bi k} - \cancel{\frac{q}{c} \tau_{\bi{k}} \left( \bi{v_k} \times \bi{B} \right) \cdot \bi{v_k} \frac{\partial g_{\bi{k}}}{\partial \tilde{\varepsilon}_{\bi k}}}, 
\end{equation}
which becomes consistent with \eref{eq:gk_isotropic} by restoring the vanishing term in \eref{eq:gk_magneto_iso_ansatz} with $\nabla_{\bi{k}}=\nabla_{\bi{n}}+\hbar \bi{v_k} \frac{\partial}{\partial \tilde{\varepsilon}_{\bi{k}}}$. 

\section{Consistency between sections \ref{sec:boltzmann_iso} and \ref{sec:boltzmann_aniso}}
\label{sec:sec12_consistency}
In this section, the consistency between sections \ref{sec:boltzmann_iso} and \ref{sec:boltzmann_aniso} is verified by showing that the result in \sref{sec:boltzmann_iso} can also be derived using the ansatz in \eref{eq:gk_magneto_aniso}.

In isotropic electron gas systems, $\bi{L_k}$ obtained from the ansatz in sections \ref{sec:boltzmann_iso} and \ref{sec:boltzmann_aniso} are given by $\bi{L_k} = \tau_{\bi{k}} \bi{v_k} \mathds{N}_{\bi{k}}$ and $\bi{L_k} = \tau_{\bi{k}} \bi{v_k} \mathds{P}_{\bi{k}}$, respectively. Therefore, the consistency can be shown by verifying $\bi{v_k} \mathds{P}_{\bi{k}} = \bi{v_k} \mathds{N}_{\bi{k}}$ in isotropic electron gas systems.  

Using \eref{eq:mu_n_def}, $\bi{l_k} = \bi{v_k} \tau_{\bi{k}} = (\hbar\tau_{\bi{k}} / m) \bi{k}$, and $\nabla_{\bi{n}} = \nabla_{\bi{k}} - \hbar \bi{v_k} \frac{\partial}{\partial \tilde{\varepsilon}_{\bi{k}}}$, we obtain the following
\begin{equation}\label{eq:Pk_iso_1st}
	\left( \mathds{F} \frac{\hat{\bmu}_{\bi n}}{c} \right) \mathds{1} = \frac{\mu_{\bi{k}}}{c} \mathds{F} - \frac{q\tau_{\bi{k}}}{c} \mathds{F} \mathds{V}_{\bi{k}},
\end{equation}
where $\mathds{V}_{\bi{k}}$ is defined by
\begin{equation}
	\mathds{V}_{\bi{k}}^{(ij)} \equiv v_{\bi{k}}^{(i)} \frac{\partial v_{\bi{k}}^{(j)}}{\partial \tilde{\varepsilon}_{\bi{k}}}.
\end{equation}
Note that $\tau_{\bi{k}}$ depends on $\bi{k}$ only through $\tilde{\varepsilon}_{\bi{k}}$. Then, using $\bi{v_k} \mathds{F} \mathds{V}_{\bi{k}} = 0$, we have
\begin{equation}\label{eq:Pk_iso_2nd}
	\left( \mathds{F} \frac{\hat{\bmu}_{\bi n}}{c} \right)^2 \mathds{1} = \left( \frac{\mu_{\bi{k}}}{c} \mathds{F} \right)^2 - \frac{q\tau_{\bi{k}}}{c} \mathds{F} \mathds{V}_{\bi{k}} \left( \frac{\mu_{\bi{k}}}{c} \mathds{F} \right).
\end{equation}
Similarly repeating the process, the following is obtained
\begin{equation}\label{eq:Pk_iso_nth}
	\left( \mathds{F} \frac{\hat{\bmu}_{\bi n}}{c} \right)^N \mathds{1} = \left( \frac{\mu_{\bi{k}}}{c} \mathds{F} \right)^N - \frac{q\tau_{\bi{k}}}{c} \mathds{F} \mathds{V}_{\bi{k}} \left( \frac{\mu_{\bi{k}}}{c} \mathds{F} \right)^{N-1}
\end{equation}
for $N \geq 1$. Substituting \eref{eq:Pk_iso_nth} to \eref{eq:Pk_alter},
\begin{equation}
	\mathds{P}_{\bi{k}} = \left( \mathds{1} - \frac{q\tau_{\bi{k}}}{c} \mathds{F} \mathds{V}_{\bi{k}} \right) \mathds{N}_{\bi{k}}.
\end{equation}
From $\bi{v_k} \mathds{F} \mathds{V}_{\bi{k}} = 0$, $\bi{v_k} \mathds{P}_{\bi{k}} = \bi{v_k} \mathds{N}_{\bi{k}}$, which completes the derivation.

\section{Details of the numerical calculation in section \ref{sec:numerical}}
\label{sec:numerical_calculation_detail}
In this section, the details of the numerical calculations in \sref{sec:numerical} are presented. Here we focus on the $xx$ component of the conductivity assuming that the orbital magnetic moment, Zeeman splitting, and internode scattering are negligible, for simplicity.

In isotropic 3D Weyl semimetals, the Hamiltonian for each node with chirality $\chi = \pm 1$ is given by
\begin{equation}
	H_\chi = \chi \hbar v_0 \bsigma \cdot \bi{k},
\end{equation}
whose eigenvalues and eigenfunctions (assuming the Fermi energy on the upper band) of the Hamiltonian are given by $\varepsilon_{\bi{k}} = \hbar v_0 k_{\rm F} r$ and
\begin{equation}
	\big| u_{\bi{k}, +} \big> = \left( \matrix{\cos(\theta/2) \cr \sin(\theta/2) e^{i\phi}} \right), \qquad \big| u_{\bi{k},-} \big> = \left( \matrix{\sin(\theta/2) \cr -\cos(\theta/2) e^{i\phi}} \right),
\end{equation}
respectively, with the overlap factors (in a single node)
\begin{equation}
	F_{\bi{k^\prime}\bi{k}} = \frac{1}{2} \left[ 1 + \cos\theta \cos\theta^\prime + \sin\theta \sin\theta^\prime \cos \left( \phi - \phi^\prime \right) \right],
\end{equation}
where $(r,\theta,\phi)$ is the spherical coordinate representing $\tilde{\bi k} \equiv {\bi k} / k_{\rm F}$, where $k_{\rm F}$ is the Fermi wavevector. Thus, the Berry curvature is given by
\begin{equation}
	\label{eq:Berry_curvature}
	\bOmega_{\bi{k},\chi} = - \textrm{Im} \left[ \big< \bi{\nabla_k} u_{\bi{k},\chi} \big| \times \big| \bi{\nabla_k} u_{\bi{k},\chi} \big> \right] = - \chi\frac{\bi{k}}{2\left| \bi{k} \right|^3} = -\frac{\chi\hat{\bi r}}{2k_{\rm F}^2 r^2}.
\end{equation}
With $\bi{B} = B \hat{\bi z}$, the phase-space volume factor $D_{\bi{k},\chi}$ is given by
\begin{equation}
	D_{\bi{k},\chi} = 1 - \frac{q}{\hbar c} \left( \bOmega_{\bi{k},\chi} \cdot \bi{B} \right) = 1 + \frac{\chi b_{\rm BC} \cos\theta}{r^2} \equiv D_\chi (\theta),
\end{equation}
where $b_{\rm BC}$ is defined by \eref{eq:b_BC_def}. Ignoring the angular magnetic moment, the velocities are simply given by 
\begin{equation}
	\bi{v_k} = v_0 \sin\theta \cos\phi \hat{\bi x} + v_0 \sin\theta \sin\phi \hat{\bi y} + v_0 \cos\theta \hat{\bi z},
\end{equation}
and the modified velocities are given by $v_{\bi{k},\chi}^{{\rm mod}(i)} = v_{\bi{k}}^{(i)} / D_{\bi{k},\chi}$ for $i = x,y$, and
\begin{equation}
	v_{\bi{k},\chi}^{{\rm mod}(z)} = \frac{v_0}{D_{\bi{k},\chi}} \left( \cos\theta + \frac{\chi b_{\rm BC}}{r^2} \right).
\end{equation}
From the Fermi golden rule, we have
\begin{equation}
	W_{\bi{k^\prime}\bi{k}} = \frac{2\pi}{\hbar} n_{\rm imp} V_0^2 F_{\bi{k^\prime}\bi{k}} \delta (\varepsilon_{\bi{k}} - \varepsilon_{\bi{k^\prime}}),
\end{equation}
where $V_0$ is the impurity potential. Then \eref{eq:Lk_integral_equation} can be rewritten as
\begin{equation}
	\label{eq:integral_equation_dimensionless}
	\tilde{\bi v}_{\bi{k},\chi}^{\rm mod} + \frac{b_\mu}{r D_{\bi{k},\chi}} \frac{\partial}{\partial \phi} \tilde{\bi L}_{\bi{k}} = 3\int \frac{d\theta^\prime d\phi^\prime}{4\pi} r^2 \sin\theta^\prime D_{\bi{k^\prime},\chi} F_{\bi{k^\prime}\bi{k}} \left( \tilde{\bi L}_{\bi{k},\chi} - \tilde{\bi L}_{\bi{k^\prime},\chi} \right),
\end{equation}
where $\tilde{\bi v}_{\bi{k},\chi}^{\rm mod} \equiv \bi{v}_{\bi{k},\chi}^{\rm mod} / v_0$, $\tilde{\bi L}_{\bi{k},\chi} \equiv \bi{L}_{\bi{k},\chi} / v_0 \tau_{\rm tr}$, $b_\mu$ is defined by \eref{eq:b_mu_def}, and $\tau_{\rm tr}$ is the relaxation time in the absence of the magnetic field at the Fermi energy given by
\begin{equation}
	\frac{1}{\tau_{\rm tr}} \equiv \frac{1}{3} \frac{2\pi}{\hbar} n_{\rm imp} V_{\rm intra}^2 \rho_{\rm F} = \frac{n_{\rm imp} V_{\rm intra}^2 k_{\rm F}^2}{3 \pi \hbar^2 v_0},
\end{equation}
where $\rho_{\rm F}$ is the density of states at the Fermi energy per degeneracy. By the symmetry of the system, each component of $\tilde{\bi L}_{\bi{k},\chi}$ should take the form
\begin{equation} \label{eq:L_tilde_form}
	\tilde{\bi L}_{\bi{k},\chi} = \tilde{L}_{\perp,\chi} (\theta) \big\{ \cos \left[ \phi - \Theta_\chi (\theta) \right] \hat{\bi x} + \sin \left[ \phi - \Theta_\chi (\theta) \right] \hat{\bi y} \big\} + \tilde{L}_{\parallel,\chi} (\theta) \hat{\bi z}.
\end{equation}
Here, $\Theta_\chi (\theta)$ describes the rotational transform induced by the magnetic field. 

Then, the $x$ component of \eref{eq:integral_equation_dimensionless} can be rewritten as
\begin{eqnarray}
	\label{eq:integral_equation_x_comp}
	&& J_{\perp,\chi} (\theta) \cos\phi - S_\chi (\theta) \tilde{L}_{\perp,\chi} (\theta) \big[ \sin\phi \cos\Theta_\chi (\theta) - \sin\Theta_\chi (\theta) \cos\phi \big] \nonumber \\ 
	&& = C_\chi (\theta) \tilde{L}_{\perp,\chi} (\theta) \big[ \cos\phi \cos\Theta_\chi (\theta) + \sin\phi \sin\Theta_\chi (\theta) \big] \nonumber \\ 
	&&  \quad - \sin\theta \left( \alpha_{\perp,\chi} \cos\phi + \beta_{\perp,\chi} \sin\phi \right), 
\end{eqnarray}
where $J_{\perp,\chi} (\theta) \equiv \sin\theta / D_\chi (\theta)$, $S_\chi (\theta) \equiv b_\mu / (rD_\chi (\theta))$,
\begin{equation}
	C_\chi (\theta) = \frac{3}{2} \left( r^2 + \frac{\chi b_{\rm BC}}{3} \cos\theta \right),
\end{equation}
and
\begin{equation} \label{eq:alpha_beta_perp}
	\left( \matrix{\alpha_{\perp,\chi} \cr \beta_{\perp,\chi}} \right) \equiv \frac{3r^2}{8} \int_0^\pi d\theta \sin^2 \theta D_\chi (\theta) \left( \matrix{\tilde{X}_\chi (\theta) \cr \tilde{Y}_\chi (\theta)} \right).
\end{equation}

Here, $\tilde{X}_\chi (\theta) \equiv \tilde{L}_{\perp,\chi} (\theta) \cos\Theta_\chi (\theta)$ and $\tilde{Y}_\chi (\theta) \equiv \tilde{L}_{\perp,\chi} (\theta) \sin\Theta_\chi (\theta)$. Comparing both sides of \eref{eq:integral_equation_x_comp}, we have
\begin{equation} \label{eq:Delta_perp_consistent}
	\left( \matrix{J_{\perp,\chi} (\theta) + \alpha_{\perp,\chi} \sin\theta \cr \beta_{\perp,\chi} \sin\theta} \right) = \left( \matrix{C_\chi (\theta) & -S_\chi (\theta) \cr S_\chi (\theta) & C_\chi (\theta)} \right) \left( \matrix{\tilde{X}_\chi (\theta) \cr \tilde{Y}_\chi (\theta)} \right).
\end{equation}
Inserting \eref{eq:Delta_perp_consistent} to \eref{eq:alpha_beta_perp},
\begin{equation} \label{eq:alpha_beta_perp_eqn}
	\left( \matrix{1 - u_{\perp,\chi} & - v_{\perp,\chi} \cr v_{\perp,\chi} & 1 - u_{\perp,\chi}} \right) \left( \matrix{\alpha_{\perp,\chi} \cr \beta_{\perp,\chi}} \right) = \left( \matrix{m_{\perp,\chi} \cr n_{\perp,\chi}} \right),
\end{equation}
where
\begin{equation}
	\left( \matrix{m_{\perp,\chi} \cr n_{\perp,\chi}} \right) = \frac{3r^2}{8} \int_0^\pi d\theta \sin^2 \theta \frac{D_\chi (\theta) J_{\perp,\chi} (\theta)}{C_\chi^2 (\theta) + S_\chi^2 (\theta)} \left( \matrix{C_\chi (\theta) \cr S_\chi (\theta)} \right),
\end{equation}
and
\begin{equation}
	\left( \matrix{u_{\perp,\chi} \cr v_{\perp,\chi}} \right) = \frac{3r^2}{8} \int_0^\pi d\theta \sin^3 \theta \frac{ D_\chi (\theta)}{C_\chi^2 (\theta) + S_\chi^2 (\theta)} \left( \matrix{C_\chi (\theta) \cr S_\chi (\theta)} \right).
\end{equation}
Solving \eref{eq:alpha_beta_perp_eqn}, we obtain $\tilde{X}_\chi (\theta)$ and $\tilde{Y}_\chi (\theta)$ through \eref{eq:Delta_perp_consistent}, and thus $\tilde{L}_{\perp,\chi} (\theta)$ and $\Theta_\chi (\theta)$.

\begin{figure}[h]
	\includegraphics[width=0.7\linewidth]{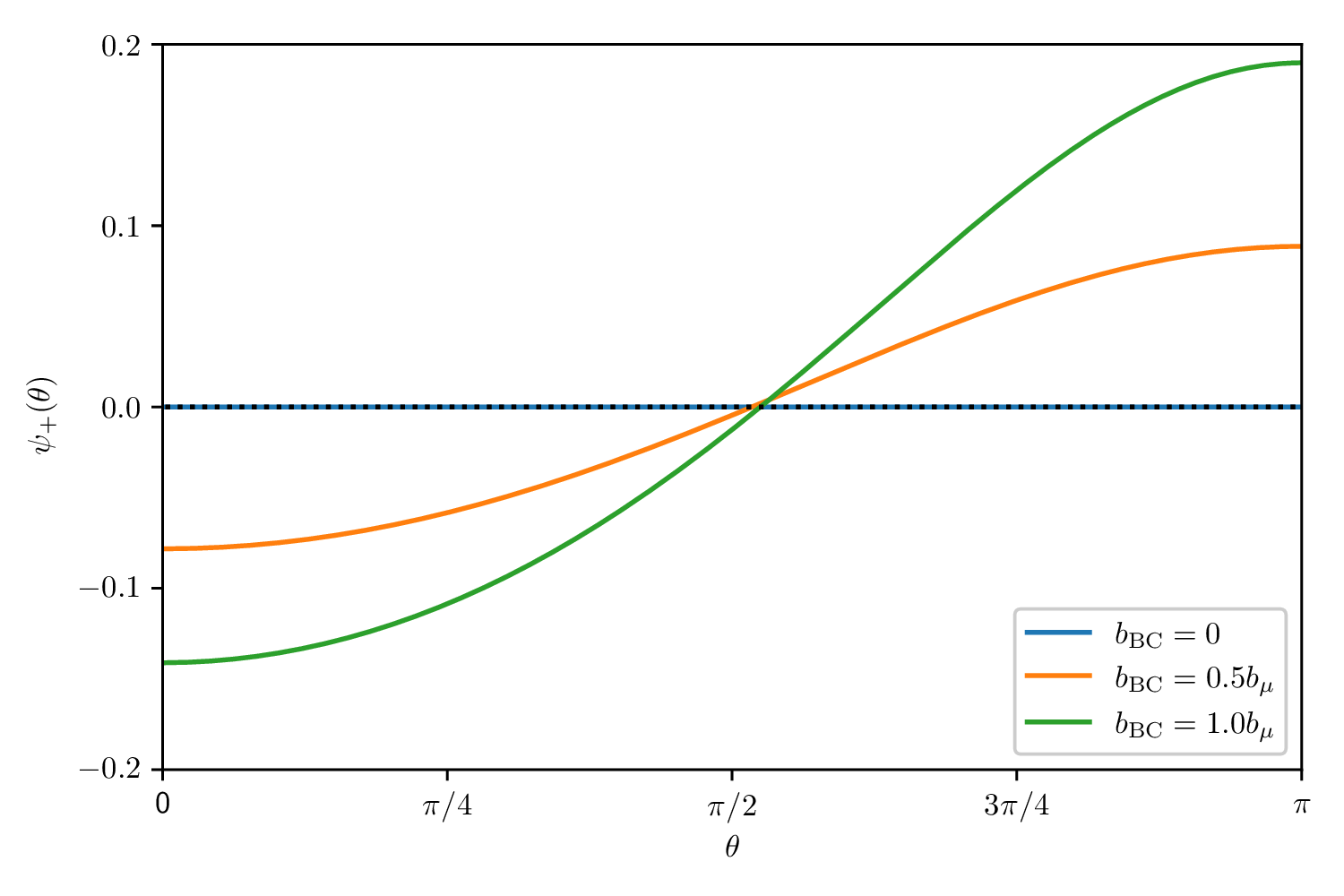}
	\centering
	\caption{
		$\psi_+ (\theta)$ as a function of $\theta$ in the $\chi=+1$ node for $b_{\rm BC} / b_{\mu}$ = 0, 0.5, and 1. Here we set $b_\mu = 0.8$.
	} 
	\label{fig:tau_xy_plot_appendix}
\end{figure}

Considering the rotational symmetry along the $z$ direction, it is reasonable to write the relaxation time on the $xy$ direction satisfying $L_{\bi{k},\chi}^{(j)} = \sum_j v_{\bi{k},\chi}^{{\rm eff}(i)} \tau^{(ij)}_{\bi{k},\chi}$ ($i,j \in \{x, y\}$) as
\begin{equation} \label{eq:tau_xy_def}
	\left( \matrix{\tau_\chi^{(xx)} (\theta) & \tau_\chi^{(xy)} (\theta) \cr \tau_\chi^{(yx)} (\theta) & \tau_\chi^{(yy)} (\theta)} \right) = \tau_{\perp,\chi} (\theta) \left( \matrix{\cos \psi_\chi (\theta) & -\sin \psi_\chi (\theta) \cr \sin \psi_\chi (\theta) & \cos \psi_\chi (\theta)} \right).
\end{equation}
Noting that the left-hand-side of \eref{eq:integral_equation_x_comp} corresponds to $\tilde{v}_{\bi{k},\chi}^{{\rm eff}(x)}$, we can easily find $\tau_{\perp,\chi} (\theta)$ and $\psi_\chi (\theta)$ from \eref{eq:L_tilde_form} as follows 
\begin{equation}
	\tilde{\tau}_{\perp,\chi} (\theta) \equiv \frac{\tau_{\perp,\chi} (\theta)}{\tau_{\rm tr}} = \frac{\tilde{L}_{\perp,\chi} (\theta)}{\sqrt{[ J_{\perp,\chi} (\theta) + S_\chi (\theta) \tilde{Y}_\chi (\theta) ]^2 + S_\chi^2 (\theta) \tilde{X}_\chi^2 (\theta)}},
\end{equation}
\begin{equation}
	\psi_\chi (\theta) = \arctan \left( \frac{\tilde{Y}_\chi (\theta)}{\tilde{X}_\chi (\theta)} \right) - \arctan \left( \frac{-S_\chi (\theta) \tilde{X}_\chi (\theta)}{J_{\perp,\chi} (\theta) + S_\chi (\theta) \tilde{Y}_\chi (\theta)} \right).
\end{equation}
The angle dependence of $\tilde{\tau}_{\perp,+} (\theta)$ for $b_{\rm BC} / b_{\mu}$ = 0, 0.5, and 1 is illustrated in figure \ref{fig:tau_xy_plot} and the corresponding $\psi_+ (\theta)$ is shown in figure \ref{fig:tau_xy_plot_appendix}.

The extrinsic magnetoconductivity given by \eref{eq:conductivity_aniso} can be written at zero temperature as follows
\begin{equation}
	\tilde{\sigma}_{ij}^{\rm ext} = \frac{3}{8\pi} \sum_\chi \int d^3 \tilde{k} D_{\bi{k},\chi} \delta(r-1) \tilde{v}_{\bi{k},\chi}^{{\rm mod}(i)} \tilde{L}_{\bi{k},\chi}^{(j)},
\end{equation}
where $\tilde{\sigma}_{ij}^{\rm ext} \equiv \sigma_{ij}^{\rm ext} / \sigma_0$ and $\sigma_0 \equiv g_{\rm n} g_{\rm s} q^2 \rho_{\rm F} v_0^2 \tau_{\rm tr} / 3$ is the conductivity in the absence of the magnetic field. Here, $g_{\rm n} = 2$ is the degeneracy for the nodes. Thus, the $xx$ component of the conductivity is given by
\begin{equation}
	\tilde{\sigma}_{xx}^{\rm ext} = \frac{3}{8} \sum_\chi \int d\theta \sin^2 \theta \tilde{X}_{\perp,\chi} (\theta),
\end{equation}
with $r=1$. Note that $\tilde{\sigma}_{xx}^{\rm ext}$ as a function of $b_{\rm BC}$ for $b_\mu = 0.8$ is illustrated in figure \ref{fig:sigma_xx_plot}.


\section*{References}
\begin{thebibliography}{999}
	
	\bibitem{Klitzing1980}
	Klitzing K V, Dorda G and Pepper M 1980 New Method for High-Accuracy
	Determination of the Fine-Structure Constant Based on Quantized Hall
	Resistance \PRL \textbf{45} 494
	
	\bibitem{Kim2013}
	Kim H-J, Kim K-S, Wang J-F, Sasaki M, Satoh N, Ohnishi A, Kitaura M, Yang M and Li L 2013 Dirac versus Weyl Fermions in Topological Insulators:	Adler-Bell-Jackiw Anomaly in Transport Phenomena \PRL \textbf{111} 246603
	
	\bibitem{Kim2014}
	Kim K-S, Kim H-J and Sasaki M 2014 Boltzmann equation approach to anomalous transport in a Weyl metal \PR B \textbf{89} 195137
	
	\bibitem{Li2016}
	Li H, He H, Lu H-Z, Zhang H, Liu H, Ma R, Fan Z, Shen S-Q and
	Wang J 2016 Negative magnetoresistance in Dirac semimetal
	${\mathrm{Cd}}_{3}{\mathrm{As}}_{2}$ {\it Nat. Commun.} \textbf{7} 10301
	
	\bibitem{Zhang2016}
	Zhang C-L \etal 2016 Signatures of the Adler-Bell-Jackiw chiral anomaly in a Weyl fermion semimetal {\it Nat. Commun.} \textbf{7} 10735
	
	\bibitem{Huang2015}
	Huang X \etal 2015 Observation of the Chiral-Anomaly-Induced Negative Magnetoresistance in 3D Weyl Semimetal TaAs \PR X \textbf{5} 031023
	
	\bibitem{Xiong2015}
	Xiong J, Kushwaha S K, Liang T, Krizan J W, Hirschberger M, Wang W, Cava R J and Ong N P 2015 Evidence for the chiral anomaly in the Dirac semimetal ${\mathrm{Na}}_{2}{\mathrm{Bi}}$ {\it Science} \textbf{350} 413
	
	\bibitem{Li2015}
	Li C-Z, Wang L-X, Liu H, Wang J, Liao Z-M and Yu D-P 2015 Giant negative magnetoresistance induced by the chiral anomaly in individual ${\mathrm{Cd}}_{3}{\mathrm {As}}_{2}$ nanowires {\it Nat. Commun.} \textbf{6} 10137
	
	\bibitem{Zhang2017}
	Zhang C \etal 2017 Room-temperature chiral charge pumping in Dirac semimetals {\it Nat. Commun.} \textbf{8} 13741
	
	\bibitem{Li2016a}
	Li Q, Kharzeev D E, Zhang C, Huang Y, Pletikosić I, Fedorov A V, Zhong R D, Schneeloch J A, Gu G D and Valla T 2016 Chiral magnetic effect in ${\mathrm{Zr}}{\mathrm{Te}}_{5}$ {\it Nat. Phys.} \textbf{12} 550
	
	\bibitem{Arnold2016}
	Arnold F \etal 2016 Negative magnetoresistance without well-defined chirality in the Weyl semimetal TaP {\it Nat. Commun.} \textbf{7} 11615
	
	\bibitem{Yang2015}
	Yang X, Li Y, Wang Z, Zhen Y and Xu Z 2015 Observation of Negative Magnetoresistance and nontrivial $\pi$ Berrys phase in 3D Weyl semi-metal NbAs arXiv:1506.02283
	
	\bibitem{Yang2015a}
	Yang X, Liu Y, Wang Z, Zheng Y and Xu Z 2015 Chiral anomaly induced negative magnetoresistance in topological Weyl semimetal NbAs arXiv:1506.03190
	
	\bibitem{Wang2016}
	Wang H \etal 2016 Chiral anomaly and ultrahigh mobility in crystalline $\mathrm{HfT}{\mathrm{e}}_{5}$ \PR B \textbf{93} 165127
	
	\bibitem{Zhang2017a}
	Zhang E \etal 2017 Tunable Positive to Negative Magnetoresistance in Atomically Thin ${\mathrm{W}}{\mathrm{Te}}_{2}$ {\it Nano Lett.} \textbf{17} 878
	
	\bibitem{Nishihaya2018}
	Nishihaya S, Uchida M, Nakazawa Y, Akiba K, Kriener M, Kozuka Y, Miyake A, Taguchi Y, Tokunaga M and Kawasaki M 2018 Negative magnetoresistance suppressed through a topological phase transition in
	$({\mathrm{Cd}}_{1\ensuremath{-}x}{\mathrm{Zn}}_{x}{)}_{3}{\mathrm{As}}_{2}$
	thin films \PR B \textbf{97} 245103
	
	\bibitem{Li2018}
	Li H, Wang H-W, He H, Wang J and Shen S-Q 2018 Giant anisotropic	magnetoresistance and planar Hall effect in the Dirac semimetal	${\mathrm{Cd}}_{3}{\mathrm{As}}_{2}$ \PR B \textbf{97} 201110(R)
	
	\bibitem{Wan2018}
	Wan B, Schindler F, Wang K, Wu K, Wan X, Neupert T and Lu H-Z 2018 Theory	for the negative longitudinal magnetoresistance in the quantum limit of	Kramers Weyl semimetals \JPCM \textbf{30}	505501
	
	\bibitem{Wang2012}
	Wang J \etal 2012 Anomalous anisotropic	magnetoresistance in topological insulator films {\it Nano Research} \textbf{5} 739
	
	\bibitem{He2013}
	He H T, Liu H C, Li B K, Guo X, Xu Z J, Xie M H and Wang J N 2013 Disorder-induced linear magnetoresistance in (221) topological insulator ${\mathrm{Bi}}_{2}{\mathrm{Se}}_{3}$ films {\it Appl. Phys. Lett.} \textbf{103}	031606
	
	\bibitem{Wiedmann2016}
	Wiedmann S \etal 2016 Anisotropic and strong negative magnetoresistance in the three-dimensional topological insulator ${\mathrm{Bi}}_{2}{\mathrm{Se}}_{3}$ \PR B \textbf{94} 081302(R)
	
	\bibitem{Wang2015}
	Wang L-X, Yan Y, Zhang L, Liao Z-M, Wu H-C and Yu D-P 2015 Zeeman effect on surface electron transport in topological insulator ${\mathrm{Bi}}_{2}{\mathrm{Se}}_{3}$ nanoribbons {\it Nanoscale} \textbf{7} 16687
	
	\bibitem{Breunig2017}
	Breunig O, Wang Z, Taskin A A, Lux J, Rosch A and Ando Y 2017 Gigantic negative magnetoresistance in the bulk of a disordered topological insulator {\it Nat. Commun.} \textbf{8} 15545
	
	\bibitem{Assaf2017}
	Assaf B A \etal 2017 Negative Longitudinal Magnetoresistance from the Anomalous $N=0$ Landau Level in Topological Materials \PRL \textbf{119}	106602
	
	\bibitem{Chen2020}
	Chen H-C, Lou Z-F, Zhou Y-X, Chen Q, Xu B-J, Chen S-J, Du J-H,
	Yang J-H, Wang H-D and Fang M-H 2020 Negative Magnetoresistance in Antiferromagnetic Topological Insulator
	${\mathrm{Eu}}{\mathrm{Sn}}_{2}{\mathrm{As}}_{2}$ \CPL
	\textbf{37} 047201
	
	\bibitem{Singh2018}
	Singh R, Gangwar V K, Daga D D, Singh A, Ghosh A K, Kumar M,
	Lakhani A, Singh R and Chatterjee S 2018 Unusual negative magnetoresistance in ${\mathrm{Bi}}_{2}{\mathrm{Se}}_{3-y} S_y$ topological insulator under perpendicular magnetic field {\it Appl. Phys. Lett.} \textbf{112} 102401
	
	\bibitem{Bhattacharyya2019}
	Bhattacharyya B, Singh B, Aloysius R P, Yadav R, Su C, Lin H, Auluck S, Gupta A, Senguttuvan T D and Husale S 2019 Spin-dependent scattering induced negative magnetoresistance in topological insulator ${\mathrm{Bi}}_{2}{\mathrm{Te}}_{3}$ nanowires {\it Sci. Rep.} \textbf{9} 7836
	
	\bibitem{Dai2017}
	Dai X, Du Z Z and Lu H-Z 2017 Negative Magnetoresistance without Chiral Anomaly in Topological Insulators \PRL \textbf{119} 166601
	
	\bibitem{Andreev2018}
	Andreev A V and Spivak B Z 2018 Longitudinal Negative Magnetoresistance and Magnetotransport Phenomena in Conventional and Topological Conductors \PRL \textbf{120} 026601
	
	\bibitem{Ishizuka2019}
	Ishizuka H and Nagaosa N 2019 Robustness of anomaly-related magnetoresistance in doped Weyl semimetals \PR B \textbf{99} 115205
	
	\bibitem{Lu2017}
	Lu H-Z and Shen S-Q 2017 Quantum transport in topological semimetals under magnetic fields {\it Front. Phys.} \textbf{12} 127201
	
	\bibitem{Behrends2017}
	Behrends J and Bardarson J H 2017 Strongly angle-dependent magnetoresistance in Weyl semimetals with long-range disorder \PR B \textbf{96} 060201(R)
	
	\bibitem{Wang2018}
	Wang H-W, Fu B and Shen S-Q 2018 Intrinsic magnetoresistance in three-dimensional Dirac materials with low carrier density \PR B \textbf{98} 081202(R)
	
	\bibitem{Fu2020}
	Fu B, Wang H-W and Shen S-Q 2020 Quantum magnetotransport in massive Dirac materials \PR B \textbf{101} 125203
	
	\bibitem{Deng2019}
	Deng M-X, Qi G Y, Ma R, Shen R, Wang R-Q, Sheng L and Xing D Y 2019 Quantum Oscillations of the Positive Longitudinal Magnetoconductivity: A Fingerprint for Identifying Weyl Semimetals \PRL \textbf{122} 036601
	
	\bibitem{Burkov2015}
	Burkov A A 2015 Chiral anomaly and transport in Weyl metals \JPCM \textbf{27} 113201
	
	\bibitem{Zyuzin2012}
	Zyuzin A A and Burkov A A 2012 Topological response in Weyl semimetals and the chiral anomaly \PR B \textbf{86} 115133
	
	\bibitem{Xiao2005}
	Xiao D, Shi J and Niu Q 2005 Berry Phase Correction to Electron Density of	States in Solids \PRL \textbf{95} 137204
	
	\bibitem{Xiao2010}
	Xiao D, Chang M-C and Niu Q 2010 Berry phase effects on electronic properties \RMP \textbf{82} 1959
	
	\bibitem{Pal2010}
	Pal H K and Maslov D L 2010 Necessary and sufficient condition for longitudinal magnetoresistance \PR B \textbf{81} 214438
	
	\bibitem{Lundgren2014}
	Lundgren R, Laurell P and Fiete G A 2014 Thermoelectric properties of Weyl and Dirac semimetals \PR B \textbf{90} 165115
	
	\bibitem{Gao2017}
	Gao Y, Yang S A and Niu Q 2017 Intrinsic relative magnetoconductivity of nonmagnetic metals \PR B \textbf{95} 165135
	
	\bibitem{Stephanov2012}
	Stephanov M A and Yin Y 2012 Chiral Kinetic Theory \PRL \textbf{109} 162001
	
	\bibitem{Son2012}
	Son D T and Yamamoto N 2012 Berry Curvature, Triangle Anomalies, and the Chiral Magnetic Effect in Fermi Liquids \PRL \textbf{109} 181602
	
	\bibitem{Son2013}
	Son D T and Spivak B Z 2013 Chiral anomaly and classical negative
	magnetoresistance of Weyl metals \PR B \textbf{88} 104412
	
	\bibitem{Zyuzin2017}
	Zyuzin V A 2017 Magnetotransport of Weyl semimetals due to the chiral anomaly \PR B \textbf{95} 245128
	
	\bibitem{Sekine2017}
	Sekine A, Culcer D and MacDonald A H 2017 Quantum kinetic theory of the chiral anomaly \PR B \textbf{96} 235134
	
	\bibitem{Olson2007}
	Olson J C and Ao P 2007 Nonequilibrium approach to Bloch-Peierls-Berry dynamics \PR B \textbf{75} 035114
	
	\bibitem{Das2019}
	Das K and Agarwal A 2019 Linear magnetochiral transport in tilted type-I and type-II Weyl semimetals \PR B \textbf{99} 085405
	
	\bibitem{Chen2016}
	Chen Q and Fiete G A 2016 Thermoelectric transport in double-Weyl semimetals \PR B \textbf{93} 155125
	
	\bibitem{Dantas2018}
	Dantas R M A, Pe{\~{n}}a-Benitez F, Roy B and Sur{\'o}wka P 2018 Magnetotransport in multi-Weyl semimetals: a kinetic theory approach \JHEP \textbf{2018} 69
	
	\bibitem{Xiao2020}
	Xiao C, Chen H, Gao Y, Xiao D, MacDonald A H and Niu Q 2020 Linear magnetoresistance induced by intra-scattering semiclassics of Bloch electrons \PR B \textbf{101} 201410(R)
	
	\bibitem{Johansson2019}
	Johansson A, Henk J and Mertig I 2019 Chiral anomaly in type-I Weyl semimetals:	Comprehensive analysis within a semiclassical Fermi surface harmonics approach \PR B \textbf{99} 075114
	
	\bibitem{Ashcroft1976}
	Ashcroft N W and Mermin N D 1976 {\it Solid State Physics} (California: Brooks Cole)
	
	\bibitem{Thonhauser2005}
	Thonhauser T, Ceresoli D, Vanderbilt D and Resta R 2005 Orbital Magnetization in Periodic Insulators \PRL \textbf{95} 137205
	
	\bibitem{Ceresoli2006}
	Ceresoli D, Thonhauser T, Vanderbilt D and Resta R 2006 Orbital magnetization in crystalline solids: Multi-band insulators, Chern insulators, and metals \PR B \textbf{74} 024408
	
	\bibitem{Souza2008}
	Souza I and Vanderbilt D 2008 Dichroic $f$-sum rule and the orbital magnetization of crystals \PR B \textbf{77} 054438
	
	\bibitem{Yao2008}
	Yao W, Xiao D and Niu Q 2008 Valley-dependent optoelectronics from inversion symmetry breaking \PR B \textbf{77} 235406
	
	\bibitem{Woo2022}
	Woo S, Min B and Min H 2022 Semiclassical magnetotransport including effects of Berry curvature and Lorentz force \PR B \textbf{105} 205126 
	
	\bibitem{Ziman2001}
	Ziman J M 2001 {\it Electrons and phonons : the theory of transport phenomena in solids} (Oxford: Oxford University Press)
	
	\bibitem{Sundaram1999}
	Sundaram G and Niu Q 1999 Wave-packet dynamics in slowly perturbed crystals: Gradient corrections and Berry-phase effects \PR B \textbf{59} 14915
	
	\bibitem{Liu2016}
	Liu Y, Low T, and Ruden P P 2016 Mobility anisotropy in monolayer black phosphorus due to scattering by charged impurities \PR B {\bf 93} 165402
	
	\bibitem{Park2017}
	Park S, Woo S, Mele E J and Min H 2017 Semiclassical Boltzmann transport theory for multi-Weyl semimetals \PR B \textbf{95} 161113(R)
	
	\bibitem{Park2019}
	Park S, Woo S and Min H 2019 Semiclassical Boltzmann transport theory of few-layer black phosphorus in various phases {\it 2D Mater.} \textbf{6} 025016
	
	\bibitem{Kim2019}
	Kim S, Woo S and Min H 2019 Vertex corrections to the dc conductivity in anisotropic multiband systems \PR B \textbf{99} 165107
	
	\bibitem{Kawamura1992}
	Kawamura T and Sarma S D 1992 Phonon-scattering-limited electron mobilities in	${\mathrm{Al}}_{\mathit{x}}$${\mathrm{Ga}}_{1\mathrm{\ensuremath{-}}\mathit{x}}$As/GaAs heterojunctions \PR B \textbf{45} 3612
	
	\bibitem{Flensberg2004}
	Bruus H and Flensberg K 2004 \textit{Many-body Quantum Theory in Condensed Matter Physics} (Oxford: Oxford University Press)
	
	\bibitem{Coleman2016} 
	Coleman P 2016 \textit{Introduction to Many-Body Physics} (Cambridge: Cambridge University Press)
\end{thebibliography}
\end{document}